# Superconductivity in the cobalt-doped V$_3$Si A15 intermetallic compound

*Lingyong Zeng*[a, #], *Huawei Zhou*[a, #], *Hong Du*[b], *Ruidan Zhong*[b], *Ruixin Guo*[c,d], *Shu Guo*[c,d], *Wanzhen Su*[a], *Kuan Li*[a], *Chao Zhang*[a], *Peifeng Yu*[a], *Huixia Luo*[a*]

[a]School of Materials Science and Engineering, State Key Laboratory of Optoelectronic Materials and Technologies, Key Lab of Polymer Composite & Functional Materials, Guangzhou Key Laboratory of Flexible Electronic Materials and Wearable Devices, Sun Yat-Sen University, No. 135, Xingang Xi Road, Guangzhou, 510275, P. R. China.

[b]Tsung-Dao Lee institute , Shanghai Jiao Tong University, Shanghai, 200240, P. R. China.

[c]Shenzhen Institute for Quantum Science and Engineering, Southern University of Science and Technology, Shenzhen 518055, China.

[d]International Quantum Academy, Shenzhen 518048, China.

[#]These authors contributed equally to this work.

[*]Corresponding author' email address: luohx7@mail.sysu.edu.cn; complete details (Telephone and Fax) (+0086)-2039386124

**Abstract**

The A15 structure of superconductors is a prototypical type-II superconductor that has generated considerable interest since the early history of superconducting materials. This paper discusses the superconducting properties of previously unreported $V_{3-x}Co_xSi$ ($0 \leq x \leq 0.30$) alloys. It is found that the lattice parameter decreases with increasing cobalt-doped content and leads to an increased residual resistivity ratio (RRR) value of the $V_{3-x}Co_xSi$ system. Meanwhile, the superconducting transition temperature ($T_c$) cobalt-doped content. Furthermore, the fitted data show that the increase of cobalt-doped content also reduces the lower/upper critical fields of the $V_{3-x}Co_xSi$ system. Type-II superconductivity is demonstrated on all $V_{3-x}Co_xSi$ samples. With higher Co-doped content, $V_{3-x}Co_xSi$ ($0.15 \leq x \leq 0.30$) alloys may have superconducting and structural phase transitions at low-temperature regions. As the electron/atom (e/a) ratio increases, the $T_c$ variation trend of $V_3Si$ is as pronounced as in crystalline alloys and monotonically follows the trend observed for amorphous superconductors.

**Introduction**

Superconductors with the cubic A15 structure, discovered mainly between the 1950s and 1970s, represent metal-based superconductors [1-5]. Until the copper-oxide superconductors were discovered in 1986, the A15 structure held the record highest superconducting transition temperature ($T_c$), around 23 K for $Nb_3Ge$ [6,7], which has since been commercialized. A15 intermetallic compounds with the general formula $A_3B$ are one of the most extensively studied materials owing to their relatively high $T_c$ values and rich physical properties. Recently, the first-principles calculations indicate that the bulk electronic band structures of $Ta_3Sb$, $Ta_3Sn$, and $Ta_3Pb$ have nontrivial band topologies, inducing the formation of topological surface states near the Fermi energy [8]. The A15 superconductors are promising candidates for the realization of topological superconductivity and the Majorana fermion.

$V_3Si$ is a well-known A15-type superconductor with the $T_c$ around 17 K and 4.7 electrons per atom (e/a) [9]. Despite its simple cubic structure at room temperature, most transport, thermodynamic and spectroscopic measurements exhibited unconventional behavior or at least some unusual features [10-15]. Such as anisotropy of the upper critical field and specific heat, de Hass-van Alphen effect, and large ratio of $T_c$ with Fermi temperature ($T_F$). Strong evidence exists that a martensitic cubic to tetragonal phase transition occurs at about 18.9 K, slightly above the $T_c$ [16-19]. Unusual softening of the acoustic phonons in $V_3Si$ has been reported [20,21]. And the inelastic x-ray scattering investigation of the lattice dynamics of $V_3Si$ indicates only a small impact of the soft phonon mode of the martensitic transition on the superconducting properties [17]. The theoretical analysis and experimental results recently present strong support for an $s^{++}$ pairing with two barely coupled gaps and weak interband coupling in the $V_3Si$ superconductor [22], which is very similar to the $MgB_2$ superconductor [23]. Besides, $V_3Si$ thin films can reach $T_c$ up to 15 K, depending on the substrate properties and the annealing temperature [24,25]. Due to the above physical properties, $V_3Si$ is still a good material platform to study physical properties, especially superconductivity. A flexible and effective method for illuminating the fundamental properties of superconductivity is the modulation of $T_c$ through carrier concentration. [26,27]. To improve V$_3$Si's superconducting properties, many chemical doping studies have been conducted. [28-30]. Nevertheless, it is still a lack of study on doping cobalt into V$_3$Si. In addition, cobalt has been widely used as a dopant to regulate the superconducting parameters or magnetic properties in iron-based superconductors (e.g., LaFe$_{1-x}$Co$_x$AsO, SmFe$_{1-x}$Co$_x$AsO, CeFe$_{1-x}$Co$_x$AsO, and Fe$_{1-x}$Co$_x$O) and cuprate superconductors (YBa$_2$Cu$_{3-x}$Co$_x$O$_{7\pm\delta}$) [31-34]. Furthermore, it is still necessary to systematically study the correlation between the valence electron count (VEC) and $T_c$ in the V$_3$Si system, especially in the region e/a > 4.7. The cobalt element ($3d^74s^2$) has 9 electrons, which is more than the 5 electrons of the vanadium element ($3d^34s^2$). The cobalt doping thus will increase the number of e/a in the V$_3$Si (e/a = 4.7) system.

In this work, magnetic cobalt ($3d^74s^2$) dopants on the vanadium site of V$_3$Si were prepared, and superconducting properties of V$_{3-x}$Co$_x$Si (0 ≤ $x$ ≤ 0.30) were systemically studied. Experimental data indicate that cobalt doping gradually suppresses $T_c$, accompanied by an increment of the e/a ratio. The fitted data show that the increase of cobalt-doped content also reduces the critical field of the V$_{3-x}$Co$_x$Si system. Type-II superconductivity is demonstrated on V$_{2.975}$Co$_{0.025}$Si alloy, whose lower critical field ($\mu_0H_{c1}(0)$) is 1.2 T and upper critical field ($\mu_0H_{c2}(0)$) is 30.0 T. With higher Co-doped content, V$_{3-x}$Co$_x$Si (0.15 ≤ $x$ ≤ 0.30) alloys may have superconducting and structural phase transitions.

**Experimental**

The polycrystalline samples V$_{3-x}$Co$_x$Si (0 ≤ $x$ ≤ 0.30) were synthesized through an arc-melting method. Stoichiometric mixtures of V (99.5%, Aladdin), Co (99.5%, Alfa Aesar), and Si (99.99%, Macklin) powders were well-ground together, pressed into a cylinder and placed into a water-cooled melting furnace. They were arc melted in an argon atmosphere and rapidly cooled. The samples were subsequently annealed at 1000 ºC under a vacuum. The weight loss during the melting process is almost negligible.

Powder x-ray diffraction (PXRD) collected the phase composition and lattice structure information from powdered samples at room temperature. We analyze the PXRD data with the Rietveld refinement model of the Fullprof suite software.

Microstructural characterization of the samples was conducted by scanning electron microscope (SEM, EVO MA10) with energy dispersive spectroscopy (EDS). Temperature-dependent resistivity (four-point probe), zero-field-cooling (ZFC) magnetic susceptibility, and heat capacity were determined through the physical property measurement system (PPMS, DynaCool, Quantum Design, Inc).

**Results and discussion**

PXRD has been recorded for identifying the composition and crystal structure of $V_{3-x}Co_xSi$ ($0 \leq x \leq 0.30$) alloys. **Fig. 1(a)** presents the detailed refinement of the typical compound $V_{2.92}Co_{0.08}Si$ at room temperature. All the diffraction peaks are crystallized in the cubic structure with *Pm-3n* space group. A good agreement is obtained between calculated and observed patterns, which is corroborated by the small $\chi^2$ (2.4214) factor. The inset in **Fig. 1(a)** displays the $V_{3-x}Co_xSi$ crystal structure based on the refinement results. The cobalt concentration is limited up to 0.3 since CoSi impurity becomes apparent as cobalt concentration is increased. As the cobalt-doped concentration increases, the (210) peak shifts to a lower Bragg angle, which is evident from the decrease in lattice constant $a$ with increasing $x$, implying the compression of the $V_3Si$ unit cell. Tiny V impurities are discovered in some cobalt-doped samples (see **Fig. 1(b)**). The relevant refined lattice parameters for $V_{3-x}Co_xSi$ ($0 \leq x \leq 0.30$) examples are shown in **Fig. 1(c)**. The lattice parameters monotonically decrease trend with enhancing cobalt-doped content, which can be attributed to the slightly smaller atomic radius of cobalt (125 pm) than vanadium (134 pm). It shows that the lattice constant $a$ reduces from 4.7243(1) Å of the pristine sample $V_3Si$ to 4.7061(2) Å of the highest doping $V_{2.7}Co_{0.3}Si$ compound. The SEM-EDS was performed to confirm the elements' distribution and the accurate ratio in the $V_{3-x}Co_xSi$ system. As shown in **Fig. 2**, a homogeneous elemental distribution was observed in the $V_{3-x}Co_xSi$ system. The cobalt is indeed in our doping alloys, and the actual cobalt ratios are very close to the designed ratios (see **Table 1**).

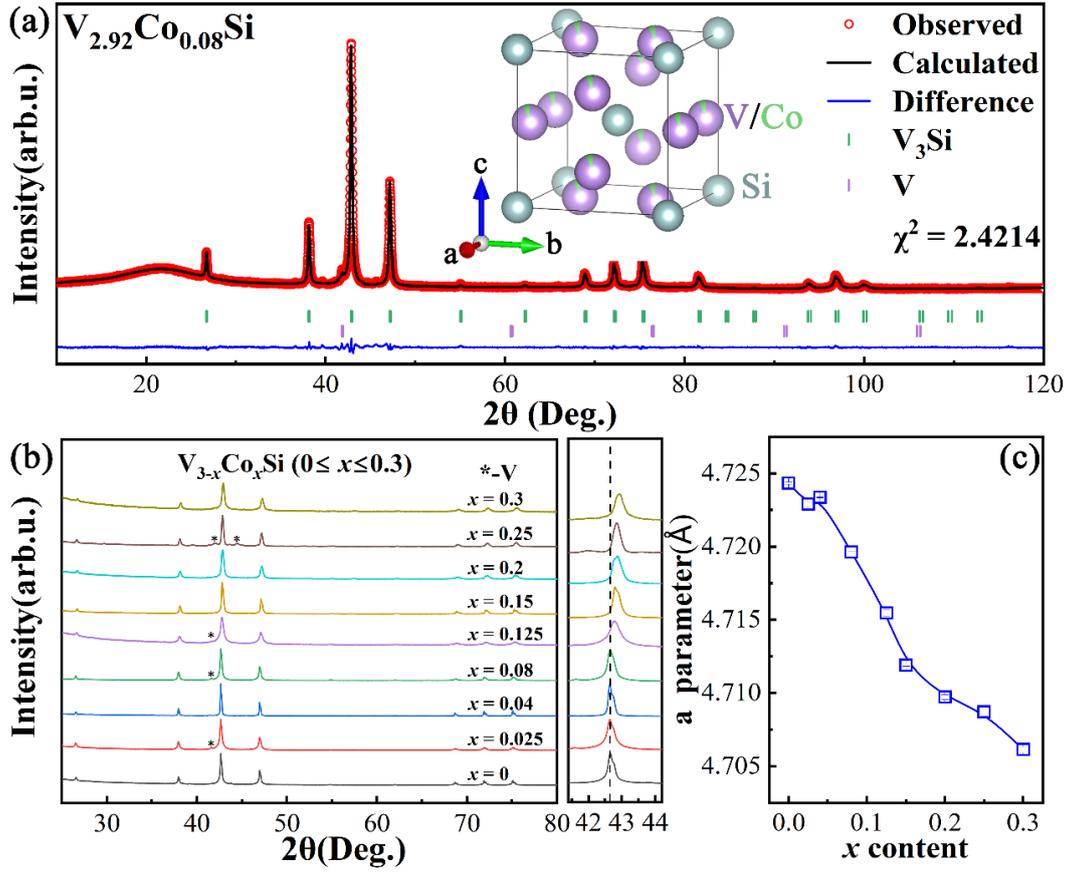

Fig. 1 (a) PXRD refinement of $V_{2.92}Co_{0.08}Si$. The inset shows the crystal structure of $V_{3-x}Co_xSi$ alloys with space group $Pm\text{-}3n$. (b) PXRD pattern of $V_{3-x}Co_xSi$ alloys. (c) Cobalt-doped content dependence of the lattice parameter $a$ of $V_{3-x}Co_xSi$ alloys.

Table 1. The element ratios of $V_{3-x}Co_xSi$ from EDS results.

| Element ratio / Sample | V | Si | Co |
|---|---|---|---|
| $V_3Si$ | 3.2 | 0.8 | 0 |
| $V_{2.975}Co_{0.025}Si$ | 3.13 | 0.85 | 0.02 |
| $V_{2.96}Co_{0.04}Si$ | 3.16 | 0.8 | 0.04 |
| $V_{2.92}Co_{0.08}Si$ | 3.07 | 0.85 | 0.08 |
| $V_{2.85}Co_{0.15}Si$ | 3 | 0.85 | 0.15 |
| $V_{2.8}Co_{0.2}Si$ | 3.01 | 0.79 | 0.2 |
| $V_{2.75}Co_{0.25}Si$ | 2.95 | 0.81 | 0.24 |

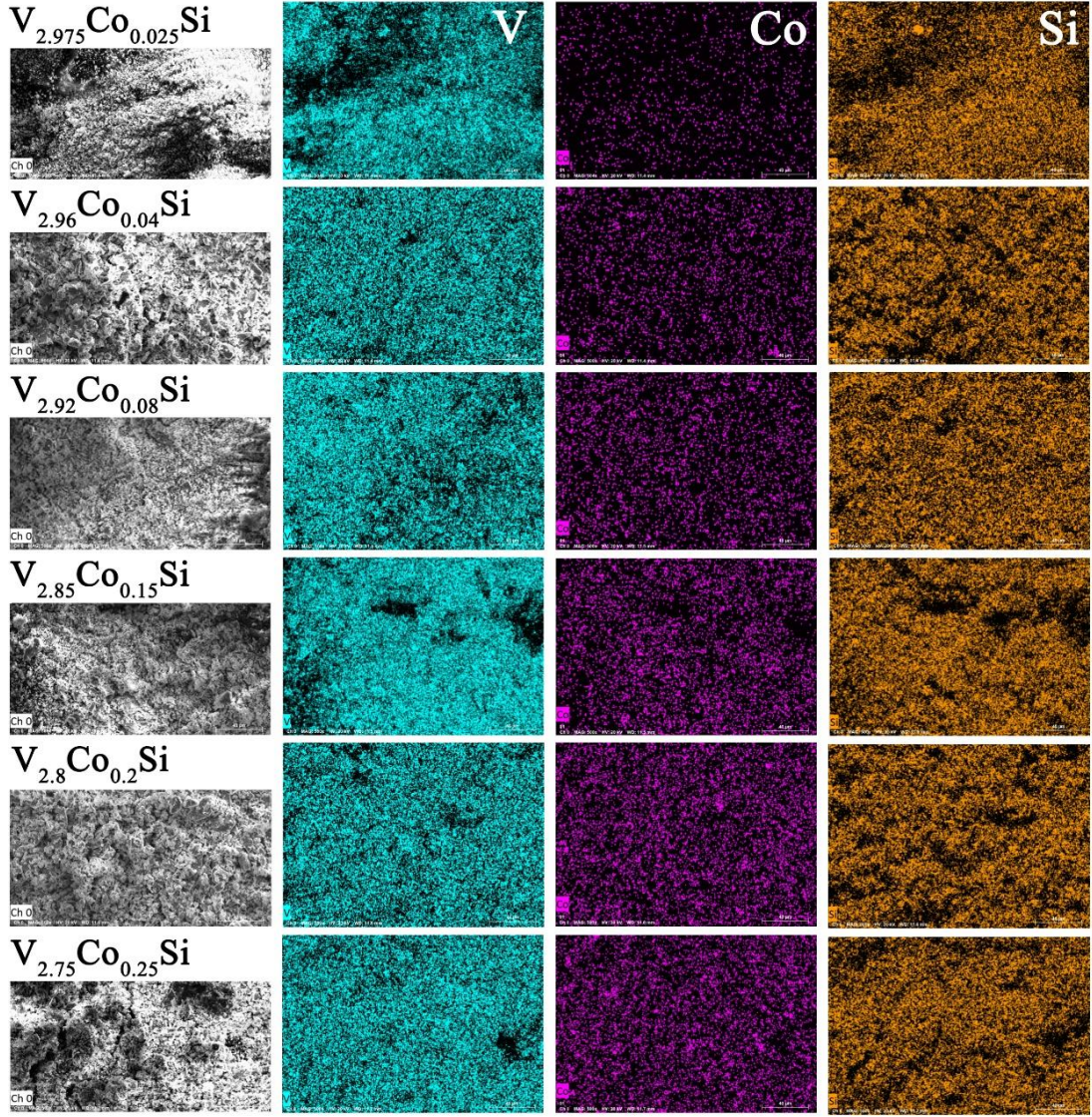

Fig. 2 SEM images and the corresponding EDS element mappings of $V_{3-x}Co_xSi$ ($0.025 \leq x \leq 0.25$).

The normalized resistivity of the $V_{3-x}Co_xSi$ ($0 \leq x \leq 0.30$) samples is shown in **Fig. 3**. The resistivity data for the $V_{3-x}Co_xSi$ system display the metallic behavior. The sharp drops can be observed in ρ(T) for the $V_{3-x}Co_xSi$ ($0 \leq x \leq 0.30$) below 18 K, which signifies the onset of the superconducting state. The resistivity trend at a temperature between 8 -18 K is shown in **Fig. 3(a)**. The $T_c$ was taken as the midpoint of the resistivity transitions. The $T_c$ of the pristine $V_3Si$ is 16.09 K, which is consistent with the previously reported value [29]. As the cobalt-doped content increases, the $T_c$ decreases, and the superconducting transition width ($\Delta T_c$) increases (see **Fig. 3(b)**). The

$T_c$ is reduced from 16.09 K for the pristine sample to 10.5 K for $V_{2.7}Co_{0.3}Si$, while the $\Delta T_c$ is increased from 0.66 K to 3.40 K. Furthermore, a small amount of cobalt-doped content causes the residual resistivity ratio (RRR = $\rho_{300K}/\rho_{18K}$) value of the $V_{3-x}Co_xSi$ system to decrease rapidly and accompanied by the slight decrease in RRR value for higher cobalt-doped concentration.

**Fig. 3(c)** shows the temperature-dependent zero-field cooling (ZFC) dc magnetic susceptibility for the $V_{3-x}Co_xSi$ ($0 \leq x \leq 0.30$) samples. Measurements were performed between 2 K and 18 K under a 20 Oe applied magnetic field. It can be seen that strong superconducting diamagnetic signals appear in all samples. The critical temperature of $T_c$ was determined as the value at the point where the linearly approximated slope crosses the zero value of normal state magnetization. To see the exact influence of the cobalt-doped content on $T_c$ obtained from the susceptibility and resistivity measurements are shown in Table 1. The $T_c$ values from magnetic and transport measurements are consistent. The $T_c$ is roughly linearly decreased with increased cobalt-doped content.

Furthermore, when $x \geq 0.15$, the change in a knee can be observed on the diamagnetic curves, and the slope at the inflection point changes significantly. **Fig. 3(d)** displays the temperature derivative of susceptibility ($d\chi/dT$) for the $V_{3-x}Co_xSi$ ($0 \leq x \leq 0.30$) samples. The red arrows indicate the $T_c$s. We can observe small humps in the temperature range of 10 - 14 K, implying the superconductivity transition. In the low temperature, sharp peaks can be observed in $V_{3-x}Co_xSi$ ($0.15 \leq x \leq 0.30$) samples, corresponding to the knee on the diamagnetic curves. We define the lowest point between the two peaks (see inset in **Fig. 3(d)**), where the slope is the smallest, and the corresponding temperature is defined as $T^*$. It can be seen that the $T^*$ decreases with increasing cobalt content. There may be a second superconducting phase or a structural phase transition at $T^*$.

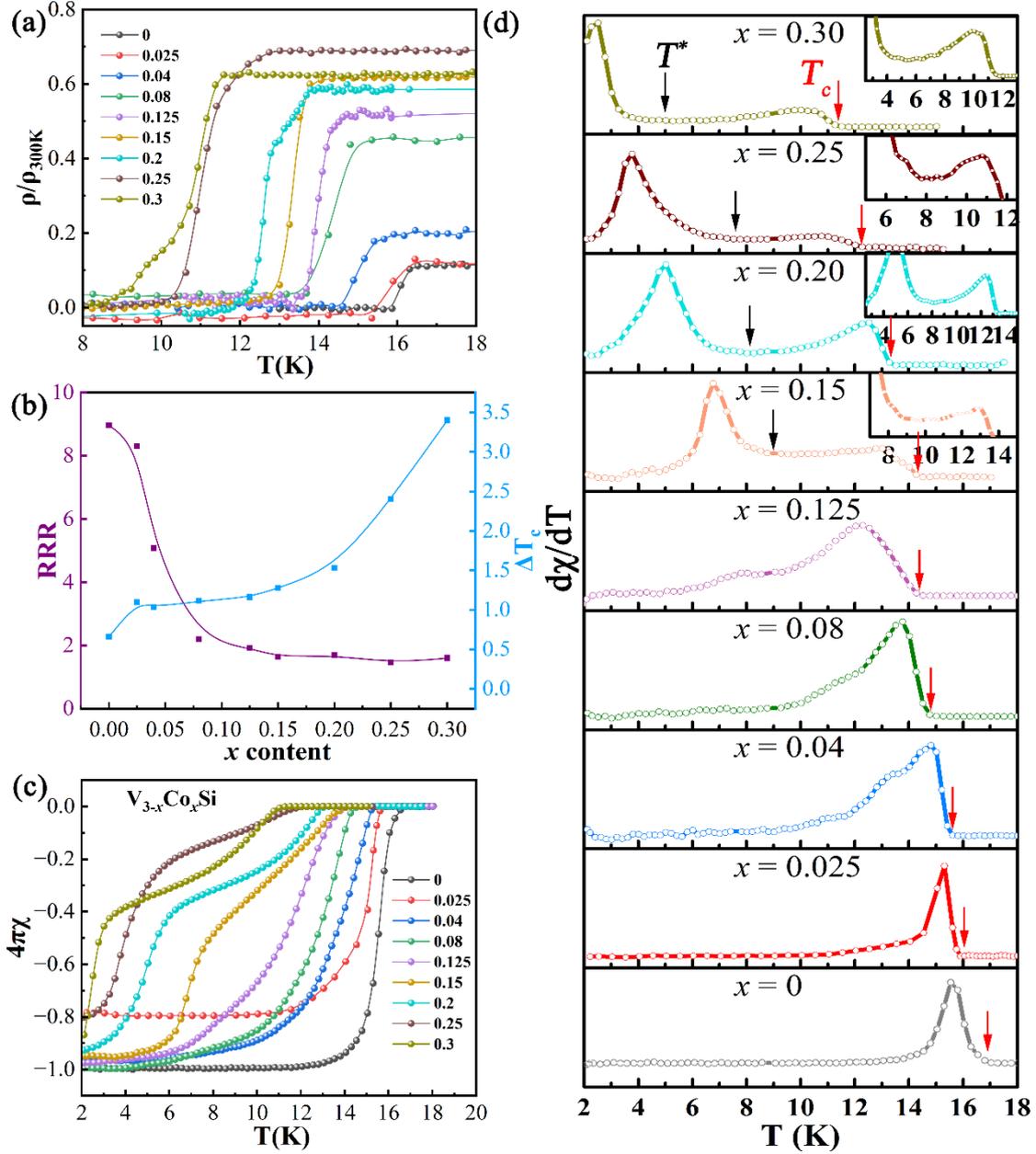

Fig. 3 (a) Normalized temperature-dependent resistivity of $V_{3-x}Co_xSi$ ($0 \leq x \leq 0.30$) alloys at the temperature range of 8-18 K. (b) cobalt content dependent on the RRR and $\Delta T_c$ values. (c) Magnetic characterization of $V_{3-x}Co_xSi$ ($0 \leq x \leq 0.30$) alloys with ZFC model. (d) the temperature derivative of susceptibility (d$\chi$/dT) for the $V_{3-x}Co_xSi$ ($0 \leq x \leq 0.30$) samples.

To further identify the lower critical field ($\mu_0H_{c1}(0)$) of $V_{3-x}Co_xSi$ ($0 \leq x \leq 0.30$) alloys, the magnetization M(H) curves at different temperatures are carried out, as shown in **Fig. 4**. The sample used for the susceptibility measurement is approximately cuboid in shape. And the magnetic field is applied in the direction of the cuboid c-axis.

The demagnetization factor (N) value is calculated from the equation $N = 4\pi\chi_v + 1$, where $\chi_v = dM/dH$ represents the slope of the linear fit. And the theoretical N value was calculated by the equation $N^{-1} = 1 + \frac{3}{4}\frac{c}{a}(1 + \frac{a}{b})$ [35], where $2a \times 2b \times 2c$ is the geometric parameters of cuboid. The geometric parameters of the susceptibility test sample are summarized in **Table 2**. At low magnetic fields, the data points have been fitted with the formula $M_{fit} = m + nH$, where m represents the intercept and n is the slope of the linear fitting. The $M-M_{fit}$ dependence of the magnetic field is present in the insets of **Fig. 4**. The $\mu_0H_{c1}(0)$ can be obtained according to the formula $\mu_0H_{c1}^*(T) = \mu_0H_{c1}^*(0)(1-(T/T_c)^2)$. Due to demagnetization, the estimated value $\mu_0H_{c1}(0)$ should be modified using the formula $\mu_0H_{c1}(0) = \mu_0H_{c1}^*(0)/(1-N)$, where the demagnetization factor N of $V_{2.975}Co_{0.025}Si$ and $V_3Si$ is 0.31 and 0.43, respectively. The N value has been used to correct the magnetic susceptibility data. The modified lower critical field $\mu_0H_{c1}(0)$ of $V_{2.975}Co_{0.025}Si$ and $V_3Si$ is 120.0 mT and 138.0 mT, respectively. All the superconducting parameters are summarized in **Table 3**. It can be seen that with the increase of cobalt content, the $\mu_0H_{c1}(0)$ also gradually decreases.

**Table 2**. The geometric parameters of the susceptibility test sample.

| Samples \ Geometric parameters | 2b (mm) | 2a (mm) | 2c (mm) | Theoretical N value | Actual N value |
|---|---|---|---|---|---|
| x = 0 | 2.55 | 1.21 | 0.98 | 0.53 | 0.43 |
| x = 0.025 | 1.58 | 1.25 | 1.20 | 0.44 | 0.31 |
| x = 0.04 | 2.01 | 1.36 | 0.99 | 0.52 | 0.31 |
| x = 0.08 | 1.75 | 1.49 | 0.70 | 0.61 | 0.51 |
| x = 0.125 | 1.35 | 0.99 | 1.20 | 0.39 | 0.36 |
| x = 0.15 | 2.01 | 1.60 | 1.49 | 0.44 | 0.51 |
| x = 0.2 | 1.38 | 1.03 | 1.16 | 0.40 | 0.40 |
| x = 0.25 | 2.18 | 1.61 | 0.53 | 0.70 | 0.58 |
| x = 0.3 | 1.98 | 1.65 | 1.24 | 0.49 | 0.51 |

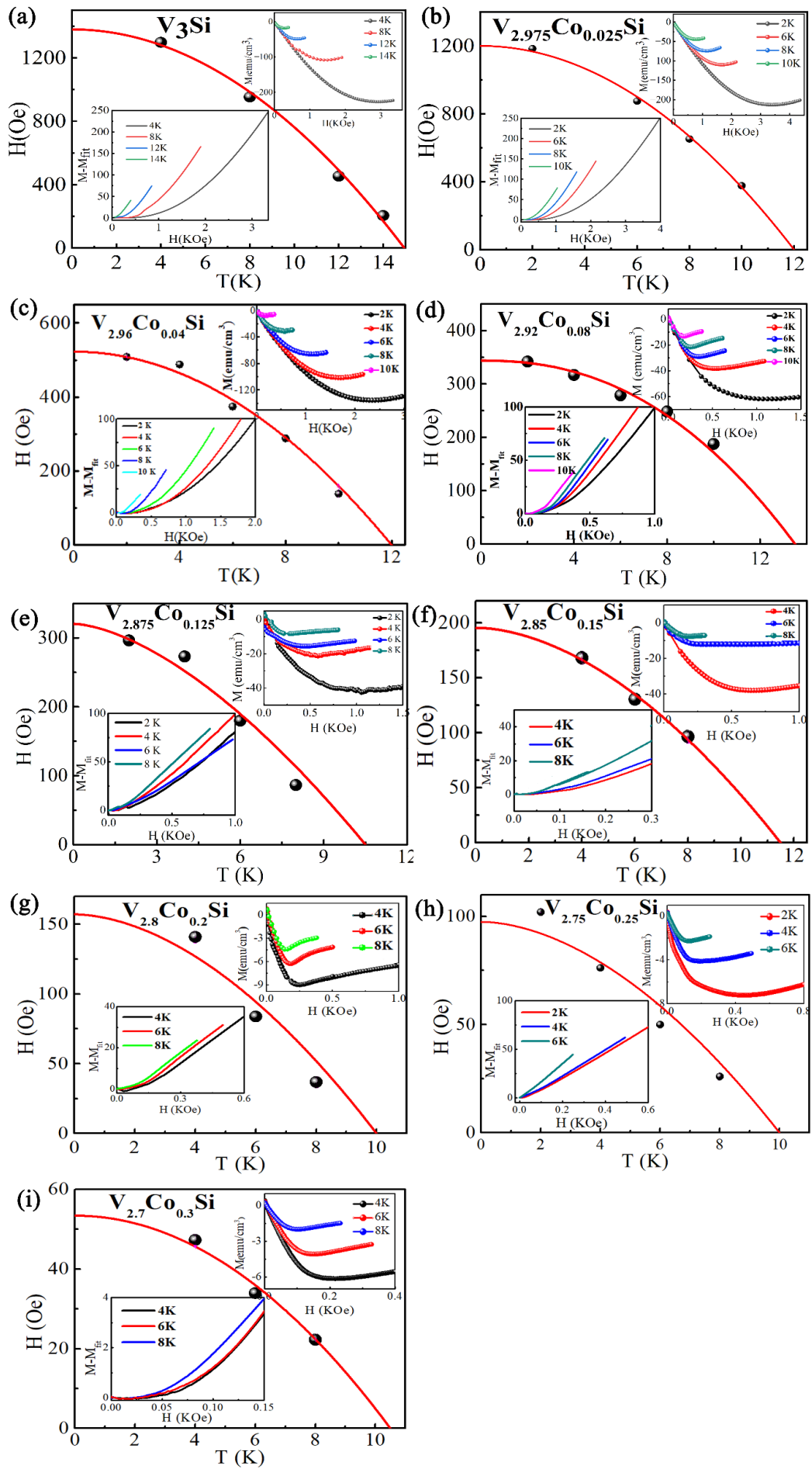

Fig. 4 The lower critical fields of $V_{3-x}Co_xSi$ ($0 \leq x \leq 0.30$) samples. The top right inset shows the M(H) curves at different temperatures. The bottom left corner inset shows the M-M$_{fit}$ as a function of the applied field.

To reveal the upper critical magnetic field ($\mu_0H_{c2}(0)$), we conducted systematic measurements around the $T_c$ under several magnetic fields of $V_{3-x}Co_xSi$ ($0 \leq x \leq 0.30$) alloys. As shown in **Fig. 5**, one can be seen that the shifting $T_c$ towards lower temperatures is visible with applying higher magnetic fields. The $\mu_0H_{c2}(0)$ is obtained by fitting the 50 % criterion of normal state resistivity values using Werthamer-Helfand-Hohenberg (WHH) and Ginzburg-Landau (GL) models. The slope ($d\mu_0H_{c2}/dT$) of $V_3Si$, $V_{2.975}Co_{0.025}Si$, and $V_{2.96}Co_{0.04}Si$ can be obtained -2.163 T/K, -2.162 T/K, and -2.161 T/K, respectively, from the linear fitting. The $\mu_0H_{c2}(0)$ can then be estimated via the WHH equation: $\mu_0H_{c2}(0) = -0.693T_c(\frac{d\mu_0H_{c2}}{dT})|_{T=T_c}$. The calculated $\mu_0H_{c2}(0)$ of the WHH model for $V_3Si$, $V_{2.975}Co_{0.025}Si$, and $V_{2.96}Co_{0.04}Si$ are 34.89 T, 34.29 T, and 32.90 T, respectively. Additionally, we fit the Pauli limiting field ($\mu_0H^P$) of $V_{3-x}Co_xSi$ ($0 \leq x \leq 0.30$) alloys by the formula $\mu_0H^P = 1.85*T_c$, the values are present in **Table 3**. The experimentally estimated $\mu_0H_{c2}(0)$ values for all samples are very close to the Pauli limiting field. Furthermore, we can also obtain the $\mu_0H_{c2}(0)$ value of 30.57 T, 30.03 T, and 28.92 T for $V_{3-x}Co_xSi$ ($x$ = 0, 0.025, 0.04) alloys fitted with GL formula: $\mu_0H_{c2}(T) = \mu_0H_{c2}(0) * \frac{1-(T/T_c)^2}{1+(T/T_c)^2}$. As displayed in **Fig. 5**, the distribution of spots obeys the function nicely. For comparison, we also calculated the $\mu_0H_{c2}(0)$ value of $V_{3-x}Co_xSi$ ($0.08 \leq x \leq 0.30$) samples. There are two superconducting transitions in the $V_{2.7}Co_{0.3}Si$ sample, possibly due to inhomogeneity. It can be seen that with the increase of cobalt content, the $\mu_0H_{c2}(0)$ also gradually decreases, except for the $V_{2.7}Co_{0.3}Si$ alloy.

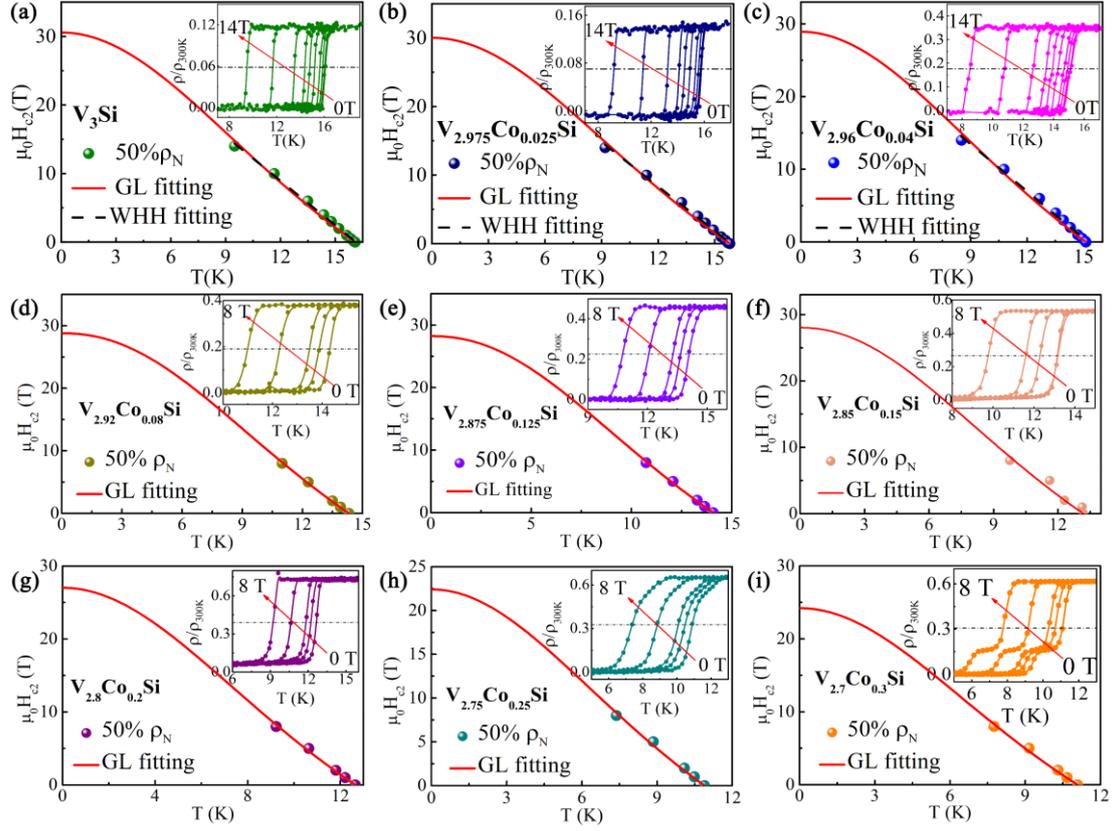

Fig. 5 The upper critical fields of $V_{3-x}Co_xSi$ ($0 \leq x \leq 0.30$) samples. The top right inset shows the low-temperature resistivity at different applied fields.

Table 3. Superconducting parameters of $V_{3-x}Co_xSi$ ($0 \leq x \leq 0.3$) alloys.

| Cobalt content ($x$) | $T_c^\rho$ (K) | $T_c^\chi$ (K) | RRR ($R_{300\,K}/R_{18\,K}$) | $\Delta T_c$ (K) | $\mu_0 H_{c1}(0)$ (mT) | $\mu_0 H_{c2}(0)$ (T) | $\mu_0 H^P$ (T) |
|---|---|---|---|---|---|---|---|
| 0 | 16.21 | 16.88 | 8.97 | 0.66 | 138.0 | 30.57 | 31.23 |
| 0.025 | 15.80 | 15.95 | 8.30 | 1.10 | 120.0 | 30.03 | 29.51 |
| 0.040 | 15.00 | 15.52 | 5.08 | 1.04 | 52.3 | 28.92 | 28.71 |
| 0.080 | 14.34 | 14.86 | 2.19 | 1.12 | 34.4 | 28.80 | 27.49 |
| 0.125 | 13.94 | 14.52 | 1.92 | 1.16 | 32.0 | 28.23 | 26.86 |
| 0.150 | 13.35 | 14.35 | 1.63 | 1.28 | 19.6 | 28.05 | 26.55 |
| 0.200 | 12.72 | 13.31 | 1.69 | 1.53 | 15.7 | 27.02 | 24.62 |
| 0.250 | 11.06 | 12.13 | 1.45 | 2.40 | 9.7 | 22.42 | 22.44 |
| 0.300 | 10.80 | 11.41 | 1.60 | 3.40 | 5.3 | 24.21 | 21.11 |

The specific heat measurement further characterizes the $V_{2.975}Co_{0.025}Si$ compound, as displayed in **Fig. 6(a)**. A sharp, specific heat jump can be observed, indicating the bulk nature of superconductivity. Meanwhile, the normal-state specific heat data have been fitted with the formula: $C_p/T = \gamma + \beta T^2$, where $\gamma$ and $\beta$ are the electronic and phonon-specific heat coefficients, respectively. The best fits give $\gamma = 0.15$ mJ/mol/K$^2$ and $\beta = 2.45 \times 10^{-4}$ mJ/mol/k$^4$. After subtracting the phonon contribution, the electronic specific heat is isolated and plotted as $C_{el}/T$ against temperature in **Fig. 6(b)**. Taking the entropy balance around the transition into account, a $T_c = 15.3$ K is obtained from the electronic-specific heat data, which agrees with the resistivity and magnetic susceptibility measurements (see **Table 3**). The electronic specific heat jump at $T_c$, $\Delta C_{el}/\gamma T_c = 1.99$, is larger than the value of 1.43 expected for the BCS weak coupling limit. Additionally, $C_{el}/T(T)$ decreases exponentially below $T_c$, suggesting an isotropic superconducting gap. The fitting performed quantitative analysis for electronic-specific heat data with a modified BCS model, the so-called α model, $C_{el}(T) = A \exp(-\Delta_0/k_B T)$, where $\Delta_0$ and $k_B$ are the superconducting gap at 0 K and Boltzmann constant, respectively. Here the coupling constant α is defined as $\alpha \equiv \Delta_0/T_c$, and the BCS weak-coupling limit $\alpha = 1.76$ (the black dash line in **Fig. 6(b)**). The obtained $\alpha = 1.20$ (solid blue line in **Fig. 5(b)**) gives an excellent fit. Furthermore, the Debye temperature ($\Theta_D$) can be calculated with the following formula: $\Theta_D = (12\pi^4 nR/5\beta)^{1/3}$, where $n = 4$ represents the number of toms per unit cell. It yields $\Theta_D = 3161$ K for $V_{2.975}Co_{0.025}Si$. Once $\Theta_D$ is calculated, the electron-phonon coupling strength $\lambda_{ep}$ can be obtained via the inverted McMillan equation: $\lambda_{ep} = \frac{1.04 + \mu^* \ln\left(\frac{\Theta_D}{1.45 T_c}\right)}{(1-0.62\mu^*)\ln\left(\frac{\Theta_D}{1.45 T_c}\right) - 1.04}$, where the Coulomb pseudopotential parameter $\mu^* = 0.13$ [26,27]. It yields $\lambda_{ep} = 0.48$ for $V_{2.975}Co_{0.025}Si$ compound.

As a comparison, we also performed the heat capacity measurement on the sample with high cobalt content. **Fig. 6(c)** displays the $C_p/T(T)$ curves of $V_{2.8}Co_{0.2}Si$ down to 2 K at 0 T and 5 T fields. However, two specific heat jumps can be observed in **Fig.**

**6(c)**, one of which has a minor anomaly in heat capacity data. The clear anomaly jump is around 12.3 K at 0 T field, which is consistent with the $T_c$ in Table 1. Meanwhile, the normal-state specific heat data also fitted with the following formula: $C_p/T = \gamma + \beta T^2$, giving $\gamma = 17.61$ mJ/mol/K$^2$ and $\beta = 0.06$ mJ/mol/k$^4$. **Fig. 6(b)** shows the $C_{el}/T(T)$ curves at a temperature between 2 - 20 K under 0 T. Using the standard equal-area entropy construction method, and we can determine $\Delta C_{el}/\gamma T_c = 1.22$, which is slightly smaller than the BCS weak coupling limit (1.43), revealing its superconducting nature. Based on the above parameters, the $\Theta_D$ and $\lambda_{ep}$ are calculated to be 506 K and 0.72, respectively. The inset in **Fig. 6(c)** displays the $C_p/T(T^2)$ data at low temperatures. Comparing the heat capacity data under the applied magnetic field of 0 T and 5 T, the external magnetic field will suppress the $T_c$, corresponding to the specific heat peak of the superconducting transition moving to a lower temperature. However, the minor anomaly in $C_p/T(T^2)$ curves around 8.5 K did not change under an external magnetic field of 5 T, which can rule out superconductivity and magnetic transitions. Interestingly, this minor anomaly of 8.5 K in the V$_{2.8}$Co$_{0.2}$Si heat capacity coincides with the $T^*$ in d$\chi$/dT curves. We speculated that a structural phase transition may have occurred at $T^*$ in the high-content cobalt-doping samples ($x \geq 0.15$). Previous studies have shown a martensitic cubic to tetragonal phase transition at the temperature of about 18.9 K in the V$_3$Si compound [16-19]. Besides, the hydrostatic pressure will gradually suppress this martensitic phase transition [36,37]. Cobalt-doping reduces the lattice parameters, which may be equivalent to applying external pressure. And the $T^*$ decreases with increasing cobalt content (see Fig. 2(d)). The V$_{2.8}$Co$_{0.2}$Si sample may have undergone a martensitic transformation at about 8.5 K. Furthermore, an in-depth study is warranted to understand the present experimental findings.

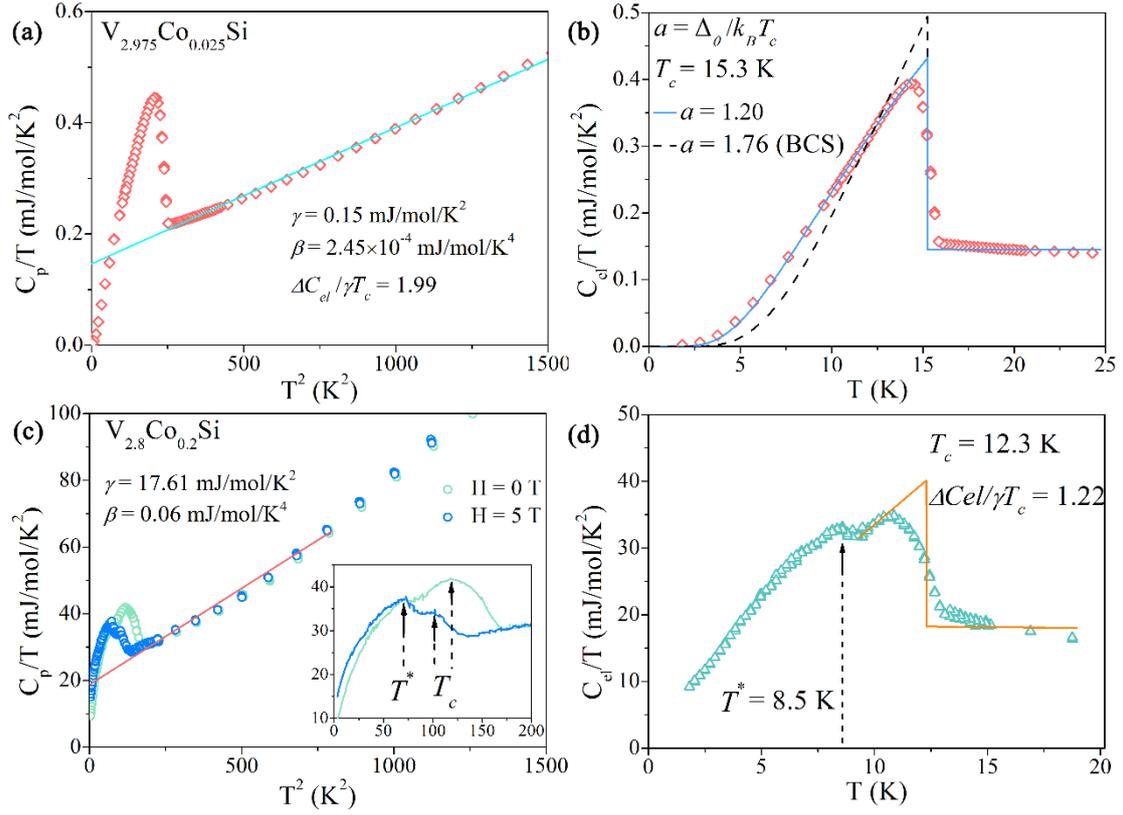

Fig. 6 (a) $C_p/T(T^2)$ curves of $V_{2.975}Co_{0.025}Si$ alloy without magnetic field. (b) Temperature dependence of the electronic specific heat $C_{el}$ of $V_{2.975}Co_{0.025}Si$ alloy. (c) $C_p/T(T^2)$ curves of $V_{2.8}Co_{0.2}Si$ alloy at 0 T and 5 T field. The inset shows the $C_p/T(T^2)$ curves at low temperatures. (d) Temperature dependence of the electronic specific heat $C_{el}$ of $V_{2.8}Co_{0.2}Si$ alloy.

**Fig. 7** plots the $T_c$ of $V_3Si_{1-x}Ga_x$ [30], $V_3Si_{1-x}Al_x$ [30], and $V_{3-x}Co_xSi$ A15 structure superconductors as a function of the electron/atom (e/a) ratio. Whether doping gallium/aluminum on silicon or cobalt on vanadium will decrease the $T_c$ of $V_3Si$, when considering the e/a ratio in the $V_3Si$ system, it can be found that the substitution of gallium/aluminum for silicon reduces the e/a ratio and the substitution of cobalt for vanadium increases the e/a ratio. We also illustrate the trend lines of the $T_c$ of transition metals and their alloys in crystalline form and as amorphous vapor-deposited films for comparison [38]. It is often referred to as the Matthias rule, which shows the $T_c$ maximum occurs at around 4.7 e/a for transition metals, even though the maximum is much broader than simple crystalline superconductors. As the e/a ratio increases, the $T_c$ variation trend of $V_3Si$ is as pronounced as in crystalline alloys and monotonically follows the trend observed for amorphous superconductors. The V$_3$Si has the maximum $T_c$ with electron count e/a = 4.7, which is the essential feature of the Matthias rule, indicating that chemical doping makes it challenging to increase the T$_c$ of V$_3$Si bulk. Furthermore, according to the rigid band model and the electronic structure calculations of V$_3$Si [39], cobalt doping at the vanadium site will increase the number of VECs in the unit cell, resulting in the increase of the Fermi energy $E_F$ and the decrease of N($E_F$).

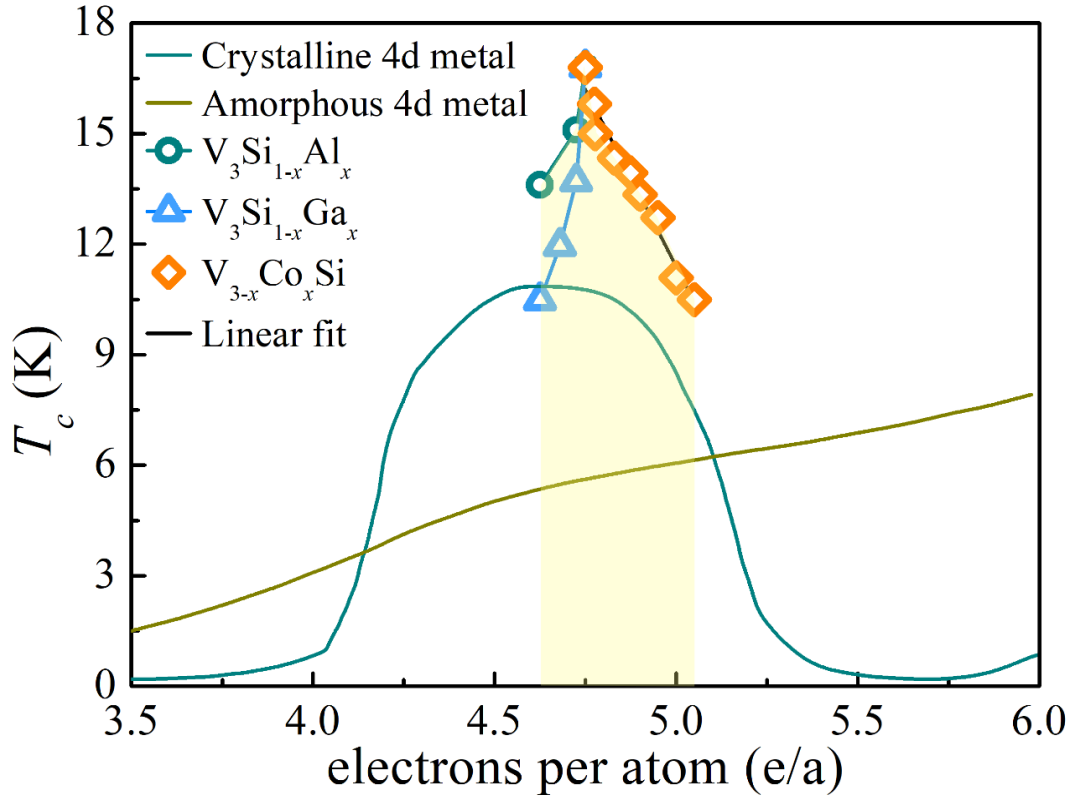

Fig. 7 valence electron count per atom dependency of the $T_c$ of V$_3$Si$_{1-x}$Al$_x$ [30], V$_3$Si$_{1-x}$Ga$_x$ [30], and V$_{3-x}$Co$_x$Si systems.

**Conclusion**

We have synthesized V$_{3-x}$Co$_x$Si alloys with $x$ up to 0.3 by an arc melting method and subsequent annealing. According to PXRD refinement data, the increase of cobalt-doped content decreases the lattice parameter. Resistivity and susceptibility measurements show an almost linear decrease in $T_c$ with increasing cobalt content in the V$_{3-x}$Co$_x$Si system. The V$_3$Si has the maximum $T_c$ with electron count e/a = 4.7, which is an essential feature of the Matthias rule. The relationship between $T_c$ and e/a ratio in the V$_{3-x}$Co$_x$Si system is almost linear. The RRR values indicate that cobalt doping introduces disorder in the V$_{3-x}$Co$_x$Si system, and the following leads to a suppression of the critical temperature $T_c$. Further, with higher Co-doped content, V$_{3-x}$Co$_x$Si (0.15 ≤ $x$ ≤ 0.30) alloys may have superconducting and structural phase transitions. The fitted data show that cobalt doping also reduces the critical field of the V$_{3-x}$Co$_x$Si system. Type-II superconductivity is demonstrated on all V$_{3-x}$Co$_x$Si samples.

**Acknowledgments**

This work is supported by the National Natural Science Foundation of China (11922415, 12274471, 22205091), Guangdong Basic and Applied Basic Research Foundation (2022A1515011168, 2019A1515011718, 2019A1515011337), the fund of State Key Laboratory of Optoelectronic Materials and Technologies (OEMT-2022-ZRC-02), the Key Research & Development Program of Guangdong Province, China (2019B110209003), and the Pearl River Scholarship Program of Guangdong Province Universities and Colleges (20191001).